\documentclass[12pt]{article}
\usepackage{graphicx}
\usepackage{enumitem}
\setlist{noitemsep,topsep=0pt,parsep=0pt,partopsep=0pt}

\newcommand\pubdate{\today}

\def\napoli{Institute of Modern Physics, \\
Chinese Academy of Sciences, Lanzhou, China}

\def\Title#1{\begin{center} {\Large #1 } \end{center}}
\def\Author#1{\begin{center}{ \sc #1} \end{center}}
\def\Address#1{\begin{center}{ \it #1} \end{center}}

\newcommand\pubblock{\rightline{\begin{tabular}{l} 
         \pubdate  \end{tabular}}}
\newenvironment{Abstract}{\begin{quotation}  }{\end{quotation}}
\newenvironment{Presented}{\begin{quotation} \begin{center} 
             PRESENTED AT\end{center}\bigskip 
      \begin{center}\begin{large}}{\end{large}\end{center} \end{quotation}}
\def\Acknowledgements{\bigskip  \bigskip \begin{center} \begin{large}
             \bf ACKNOWLEDGEMENTS \end{large}\end{center}}

\def\beq{\begin{equation}}
\def\eeq#1{\label{#1}\end{equation}}
\def\eeqn{\end{equation}}

\def\beqa{\begin{eqnarray}}
\def\eeqa#1{\label{#1}\end{eqnarray}}
\def\eeqan{\end{eqnarray}}

\let\bar=\overbar

\def\Dslash{\not{\hbox{\kern-4pt $D$}}}
\def\dslash{\not{\hbox{\kern-2pt $\del$}}}

\def\msb{{\bar{\ssstyle M \kern -1pt S}}}

\begin{document}
\begin{titlepage}
\pubblock

\vfill
\Title{Statistical Methods for the Neutrino Mass Hierarchy}
\vfill
\Author{ Emilio Ciuffoli}
\Address{\napoli}
\vfill
\begin{Abstract}
In the next decade several experiments will attempt to determine the neutrino mass hierarchy, i.e. the sign of  $\Delta m_{31}^2$. In the last years it was noticed that the two hierarchies are disjoint hypotheses and, for this reason, Wilks' theorem cannot be applied: this means that $\Delta\chi^2=\chi^2_{IH}-\chi^2_{NH}$ does not follow a one-degree-of-freedom chi-square distribution. It was proven that, under certain assumptions, it follows instead a Gaussian distribution with $\sigma=2\sqrt{\mu}$. I will present several possible definition of sensitivity and review the approaches proposed in the literature, both within the Bayesian and the frequentist framework, examining advantages and disadvantages and discussing how they should be modified if the conditions for Gaussianity are not fulfilled. I will also discuss the possibility of introducing a new pull parameter in order to avoid the issue related to the non-nested hypotheses and the differences between marginalization and minimization, showing under which conditions the two procedures yield the same $\Delta\chi^2$.
\end{Abstract}
\vfill
\begin{Presented}
 NuPhys2016, Prospects in Neutrino Physics

Barbican Centre, London, UK,  December 12--14, 2016
\end{Presented}
\vfill
\end{titlepage}
\def\thefootnote{\fnsymbol{footnote}}
\setcounter{footnote}{0}

\section{The Statistical Problem}
In the next decade several experiments will attempt to determine the neutrino mass hierarchy, {\it i.e.} the sign of $\Delta m_{31}^2$: if it is positive, the hierarchy is called normal, if negative inverted. To estimate the robustness of the mass hierarchy determination achieved by an experiment we use a test statistic, namely a function of the data whose value is related to the mass hierarchy (for example, it can be larger when the hierarchy is normal). In this work, as test statistic, we will use $\Delta\chi^2$ defined as
\begin{equation}\label{defChi}
\Delta\chi^2=\chi^2_{IH} - \chi^2_{NH}=-2\textrm{ln}\frac{P(D|IH)}{P(D|NH)}
\end{equation}
where $P(D|NH(IH))$ is the likelihood of getting the data $D$ assuming the hierarchy to be normal (inverted). This is not the only possible choice: for example, an alternative test statistic was proposed in \cite{stanco}. In general, $\chi^2$ can depend on several additional parameters (pull parameters), which are not directly related to the mass hierarchy determination: we will indicate them with $\theta$ (it could represent a a single parameter as well as a vector). There are different ways to treat these parameters, for example in the frequentist approach (which is the approach that will be used in the two examples described below in this section), they should be minimized, namely, using the value of $\theta$ which minimized the $\chi^2$ (best fit value). $\Delta\chi^2$ is now defined as
\begin{equation}\label{defChi}
\Delta\chi^2=\chi^2_{IH}(\widehat{\theta}) - \chi^2_{NH}(\widehat{\widehat{\theta}})
\end{equation}
where $\widehat{\theta}$ and $\widehat{\widehat{\theta}}$ are the best fit values for $\theta$ for each hierarchy, respectively.

Two hypotheses $H_0$ and $H_1$ are called nested if one is a particular case of the other (namely, $H_0\subset H_1$): for example, $H_0$: ``A is equal to zero'' versus $H_1$: ``A is real''. Under some very general assumptions, Wilks' theorem states that in the case of nested hypothesis $\Delta\chi^2$ follows a chi-square distribution and the confidence level with which it is possible to reject the hypothesis $H_0$, expressed in the usual form of number of Gaussian standard deviations n (number of $\sigma$'s), is equal to $\sqrt{\Delta\chi^2}$. The relation between n and the probability $p$ for $H_0$ to be true is \footnote{Here we considered the one-sided Gaussian fluctuation, for the two-sided definition $p\rightarrow 2p$ \cite{cramer}}
\begin{equation}\label{sigma}
p=\frac{1}{2}\textrm{Erfc}(n/\sqrt{2})
\end{equation}
A few years ago was noted that the two hierarchies are non-nested (or disjoint) hypotheses and, for this reason, Wilks' theorem cannot be applied. The main consequence is that the $\Delta\chi^2$ does not follow a one-degree-of freedom chi-square distribution, and the number of $\sigma$'s should not be simply estimated as $\sqrt{\Delta\chi^2}$. Indeed, in the case of the mass hierarchy we are comparing the hypothesis $H_0$: ``the sign of $\Delta m_{31}^2$ is +1'' with $H_1$: ``the sign of $\Delta m_{31}^2$ is -1''; in this case $H_0$ is not a particular case of $H_1$ and Wilks' theorem cannot be applied: to convince ourselves of this fact, we can notice that $\Delta\chi^2$ defined as Eq. (\ref{defChi}) could also be negative, if the hierarchy is inverted, while any quantity that follows a chi-square distribution must always be positive.

Under certain conditions the statistical distribution of $\Delta\chi^2$ can be described to a very good approximation by a Gaussian distribution, with $\mu=\overline{\Delta\chi^2}$ and $\sigma=2\sqrt{|\overline{\Delta\chi^2}|}$, where $\overline{\Delta\chi^2}$ is the expected value of the test statistic in question: this was first proved by Qian {\it et al.} \cite{qian}, without taking into account the eventual pull parameters ({\it ``simple vs. simple''} scenario); then this result was extended considering also pull parameters \cite{shaofeng, stat, blenFreq}, using both the Bayesian and frequentist approaches. This is not the first case in physics of non-nested hypotheses: for example Cousins {\it et al.} faced a similar problem discussing the discrimination between spin-1 and spin-2 resonances at LHC \cite{cousins}. $\overline{\Delta\chi^2}$ is equal (due to the law of large numbers) to the $\Delta\chi^2$ calculated with the Asimov data set, namely using the theoretical prediction for the expected number of events in every bin; when the $\Delta\chi^2$ follows a Gaussian distribution, the ``median experiment'' is defined as the experiment where $\Delta\chi^2=\overline{\Delta\chi^2}$; while this is always true in the Gaussian case, it is not true in general. If we define as $y_i(\theta_j)$ the expected number of events for every bin i, which in general is a function of a certain number of pull parameters $\theta_j$, the conditions that must be fulfilled in order to ensure Gaussianity are
\begin{itemize}
\item $y_i$ can be approximated as a linear function of $\theta_j$: this define a P-hyperplane in the N-dimensional space, where N is the number of bins, P is the number of pull parameters
\item  The hyperplanes for the two hypotheses are parallel around the minima
\end{itemize}
I will discuss two simplified models,  inspired by reactor and accelerator neutrino experiments, to clarify when the conditions for Gaussianity are satisfied.

\begin{figure}\begin{center}
\includegraphics[height=3.cm]{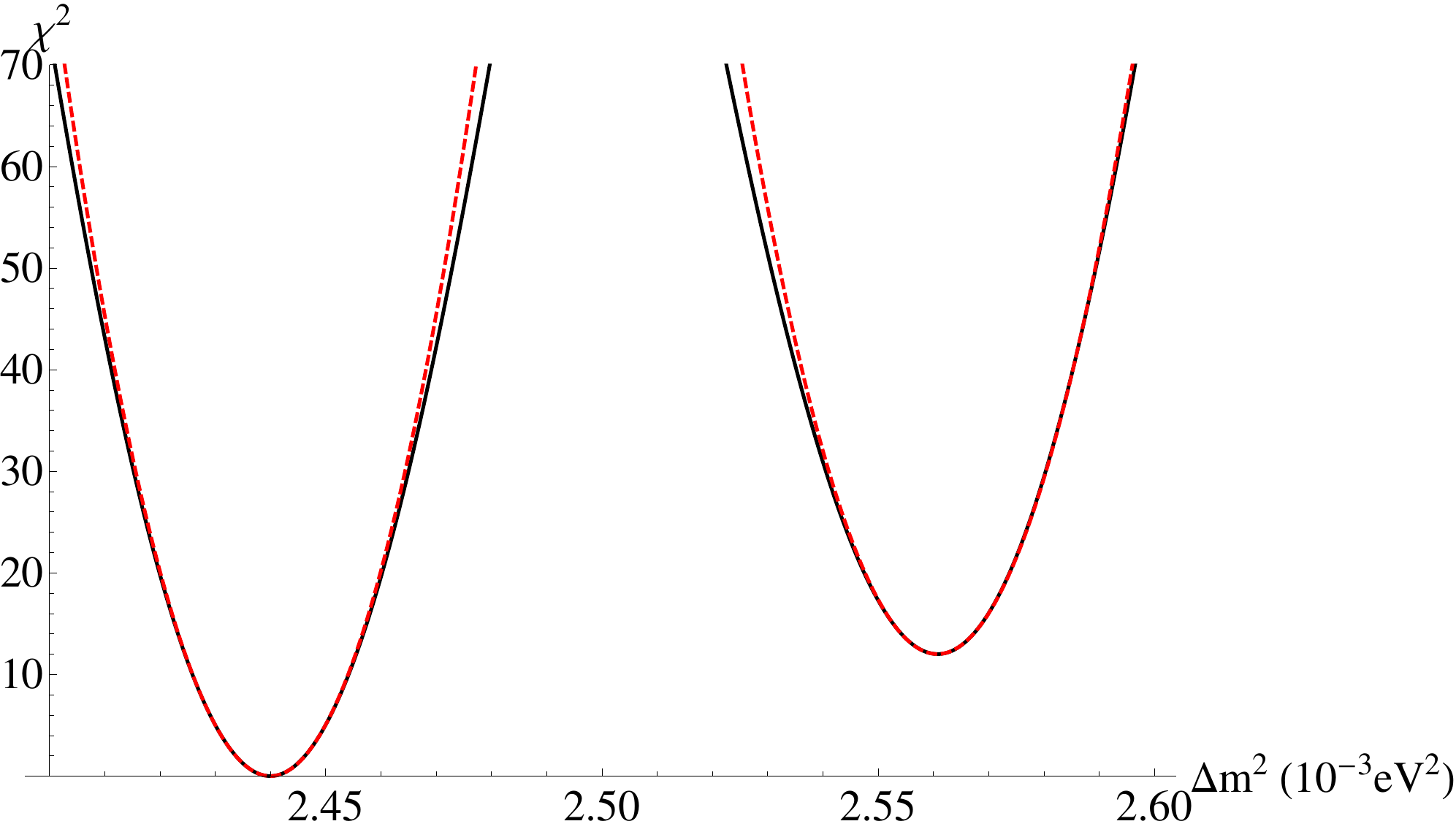}
\includegraphics[height=3.8cm]{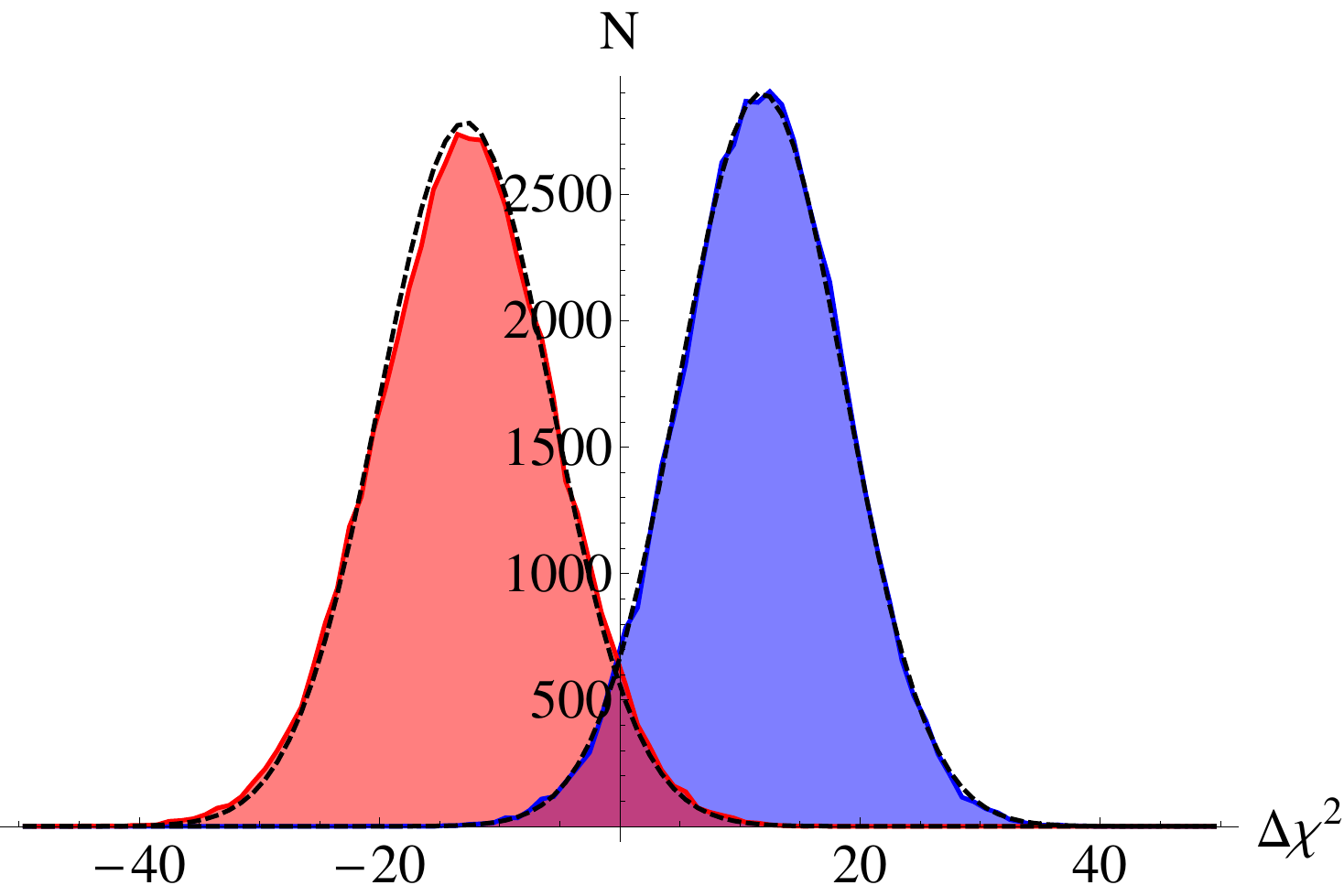}
\end{center}
\caption{\label{plotReact} Left: Asimov $\chi^2$ for normal and inverted hierarchy (black) and parabolic fit (red dashed) in reactor neutrino experiments. Right: statistical distribution of $\Delta\chi^2$}
\end{figure}

The possibility to use reactor neutrinos to determine the mass hierarchy was suggested for the first time by Petcov and Piai in 2002 \cite{petcov}. In this kind of experiment, due to the energy range of neutrinos, the matter effect are completely negligible. Vacuum oscillations depend only on the absolute values of $\Delta m^2$'s, however they obey the relation
\begin{equation}\label{deltaM}
|\Delta m_{31}^2|=|\Delta m_{32}^2|\pm|\Delta m_{21}^2|
\end{equation}
where the sign depends on the mass hierarchy. Studying the  interference between 1-2 and 1-3 oscillations it is possible to determine the mass hierarchy, but there is a strong degeneracy between a shift of $\Delta m_{31}^2$ and a change of hierarchy. We considered a model with only one pull parameter, $\Delta m_{31}^2$, ignored the background and assumed an energy resolution of $3\%/\sqrt{E}$. The baseline considered was 52 km, and the exposure 120 ktons-years. From Fig. \ref{plotReact} we can see that the Asimov $\chi^2$ is almost exactly parabolic, which is a necessarily condition for the Gaussianity, and the $\Delta\chi^2$ follows a Gaussian distribution.

\begin{figure}\begin{center}
\includegraphics[height=3.2cm]{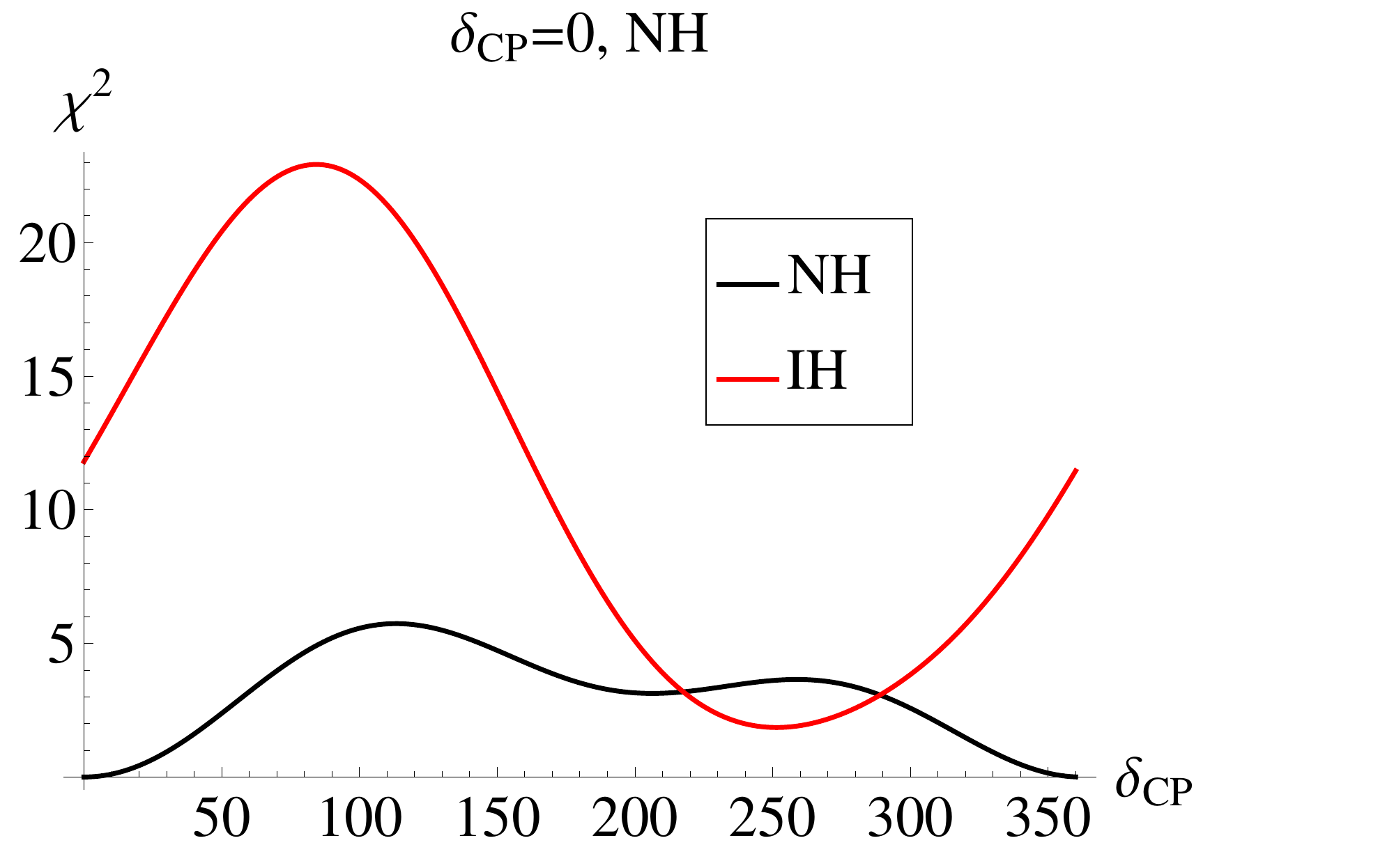}
\includegraphics[height=3.2cm]{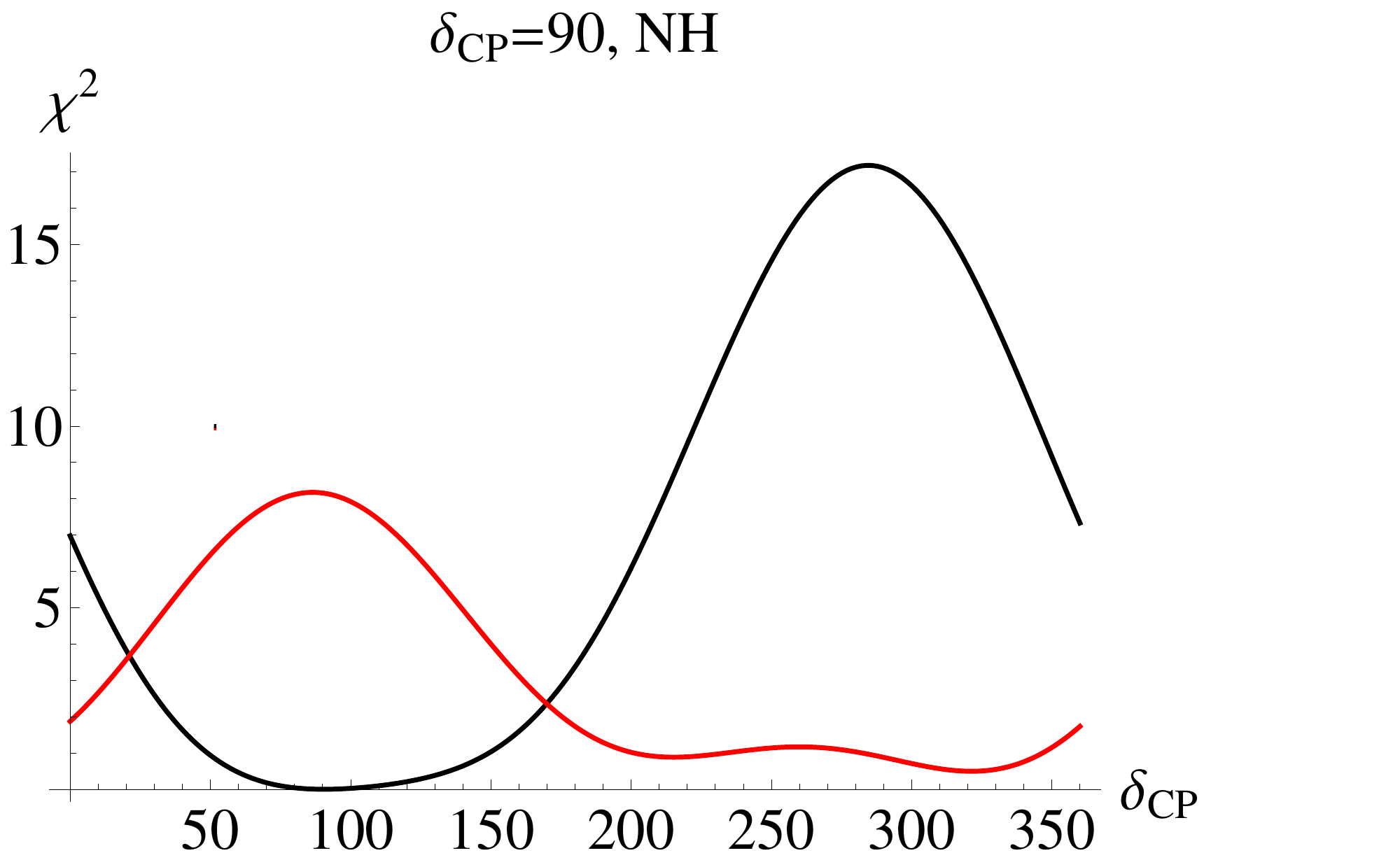}
\end{center}
\caption{\label{plotChiAcc} Asimov $\Delta\chi^2$ for different values of $\delta_{CP}$ (accelerator neutrino experiments)}
\end{figure}

\begin{figure}\begin{center}
\includegraphics[height=3.2cm]{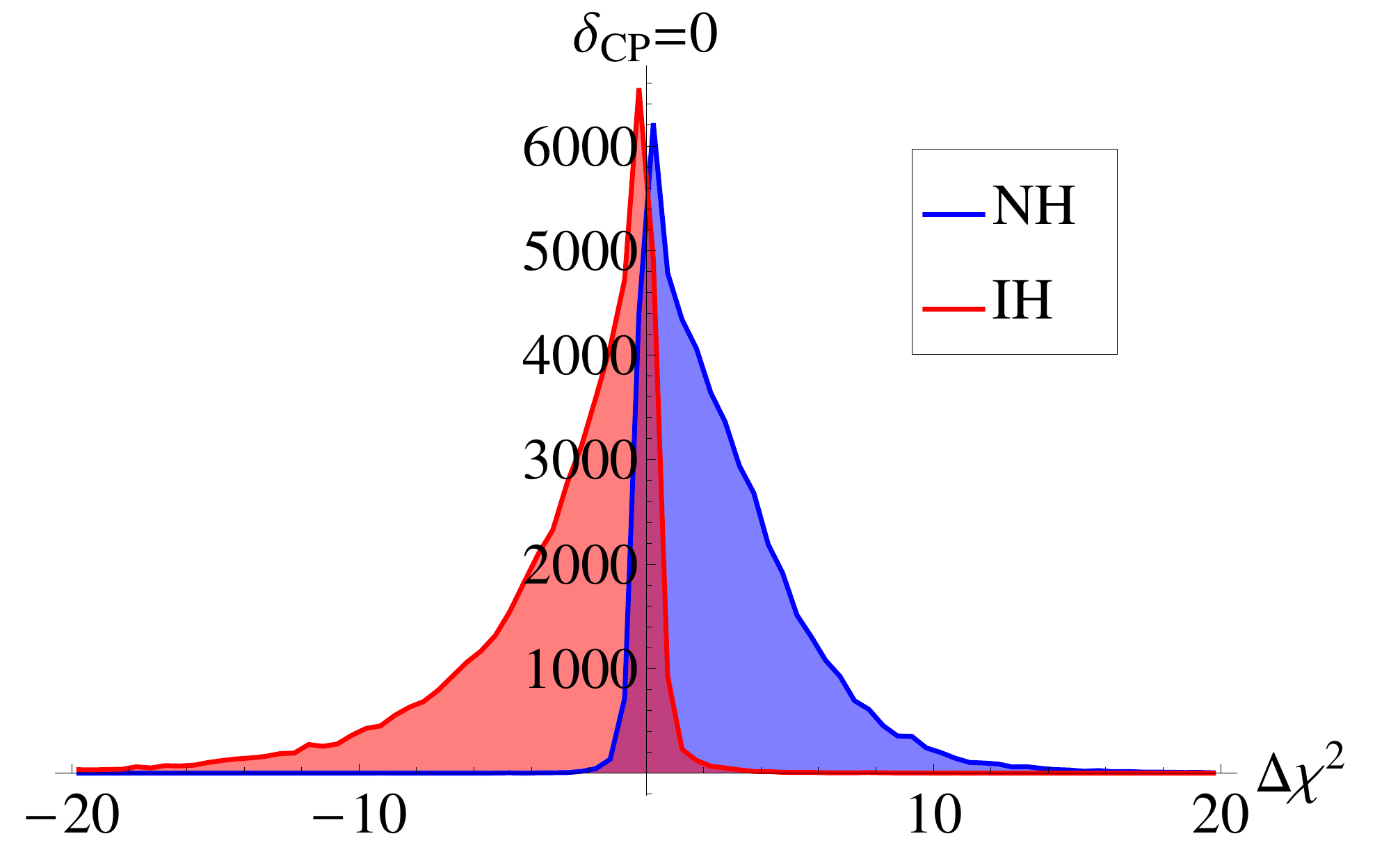}
\includegraphics[height=3.2cm]{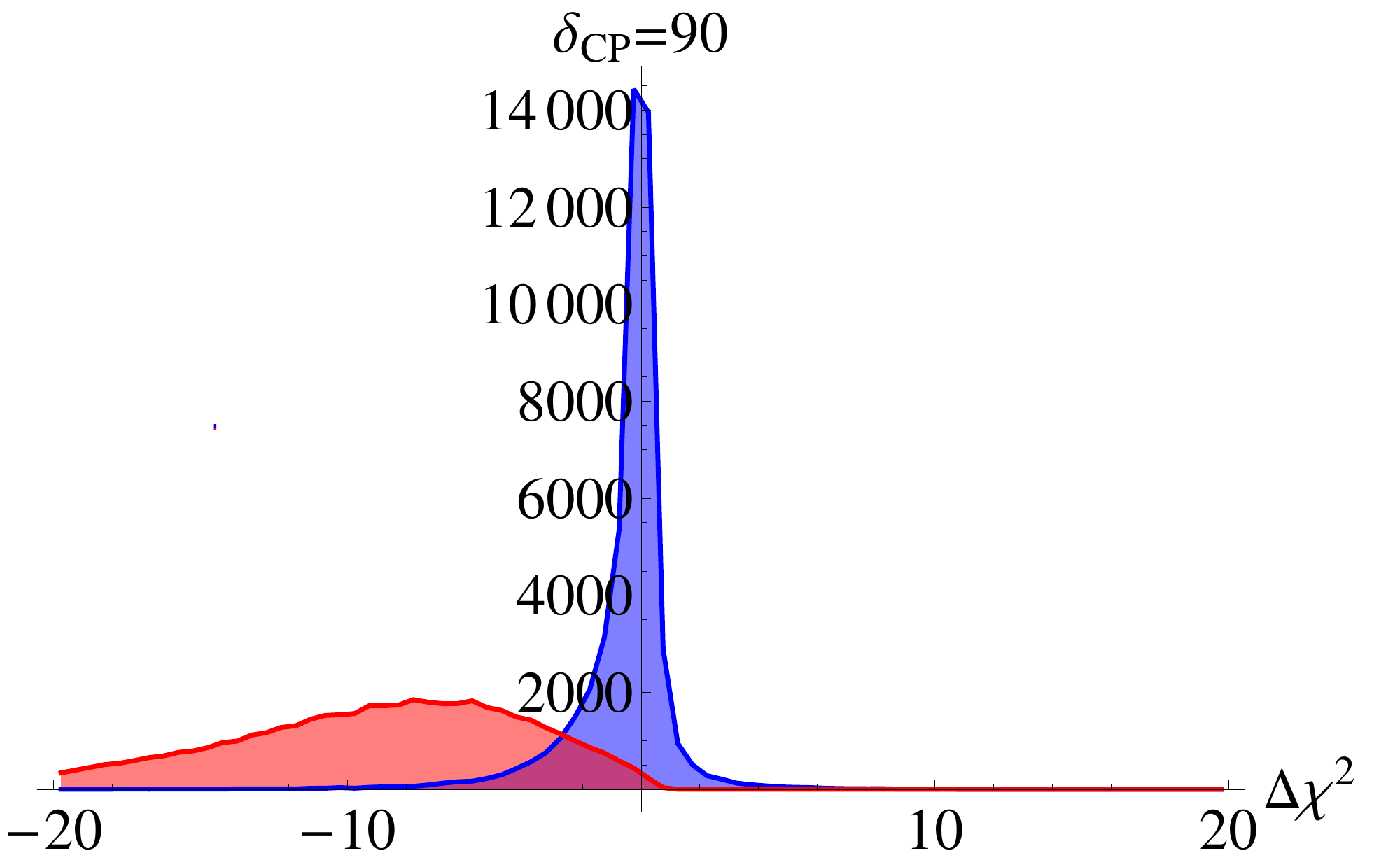}
\end{center}
\caption{\label{plotDeltaAcc} Pdf of $\Delta\chi^2$ for different values of $\delta_{CP}$ (accelerator neutrino experiments)}
\end{figure}

In accelerator neutrino experiments, instead, the mass hierarchy can be obtained by comparing the oscillation probabilities in the neutrino and antineutrino sector. While in reactor neutrino experiments we study the survival probability $P_{e\rightarrow e}$, here we observe the oscillation probability $P_{\mu\rightarrow e}$: one of the consequences is that the strongest degeneracy is now due to $\delta_{CP}$, which is only partially broken by the matter effect: in particular $P_{\mu\rightarrow e}(NH, \delta_{CP}\simeq90)\simeq P_{\mu\rightarrow e}(IH,\delta_{CP}\simeq270)$. We considered a very simplified model, with one pull parameter, $\delta_{CP}$, and where only the average oscillation probability in the neutrino and antineutrino sector was taken into account (namely, no spectral information); again all the possible sources of background were neglected. The baseline and the expected number of events were the same as a 3+3 years NO$\nu$A run \cite{nova}. In Fig. \ref{plotChiAcc} we can see that the Asimov $\chi^2$, in the case of accelerator neutrinos, is not parabolic anymore; this means that the conditions for Gaussianity are no longer fulfilled, as can be seen also from Fig. \ref{plotDeltaAcc}. The asymmetry between the probability density function (pdf) of $\Delta\chi^2$ for certain values of $\delta_{CP}$ is to be expected, and it is due to the partial degeneracy mentioned before.

%

%
%

\section{Quantify the Sensitivity to the Mass Hierarchy}
\subsection{Frequentist Approach}
We want to compare a hypothesis $H_0$ (also called ``null hypothesis'') with an alternative hypothesis (or set of alternatives) $H_1$. In order to perform a frequentist hypothesis test, we define a test statistic T and a threshold $T_c$ (let us assume that a large value of T means that $H_0$ is unlikely): if, after performing the experiment, we find a value of $T_{obs}<T_c$, $H_0$ is accepted, otherwise is rejected. In this kind of test, there are two relevant quantities:
\begin{itemize}
\item The probability $\alpha$ of rejecting $H_0$ even if it is true; $1-\alpha$ is called the confidence level (CL)
\item The probability $\beta$ of not rejecting $H_0$ even if the alternative hypothesis $H_1$ is true; $1-\beta$ is called the power of the test
\end{itemize}
It is important to underline that rejecting $H_0$ does not give, a priori, any information on $H_1$: if a hypothesis test excludes the normal hierarchy  with a certain CL, this does not tell us anything about the inverted hierarchy. In \cite{blenFreq}, the authors suggest to test both hypotheses separately, defining two threshold, $T_{c,NH}$ and $T_{c,IH}$. As test statistic we use $\Delta\chi^2$, defined as (\ref{defChi}): if our experiment gives us $\Delta\chi^2<T_{c,NH}$, the normal hierarchy is rejected, if we find  $\Delta\chi^2>T_{c,IH}$, the inverted hierarchy is rejected. One unappealing consequence of this approach is that, depending on the choices of  $T_{c,NH}$ and $T_{c,IH}$, both hierarchies can be accepted or rejected at the same time. The CL that can be achieved depends only on the values of  $T_{c,NH}$ and $T_{c,IH}$ (that must be chosen before the experiment), not on the results obtained: for this reason it may be convenient to use the frequentist hypothesis test to estimate the expected sensitivity that can be achieved in a future experiment. In \cite{blenFreq} different definitions of sensitivity are proposed:
\begin{itemize}
\item The median sensitivity is defined by choosing $T_{c,NH (IH)}=\overline{\Delta\chi^2}_{IH (NH)}$, namely the expected $\Delta\chi^2$ if the hierarchy is the opposite with respect to what we are testing. In particular, if we assume the symmetric case where $\overline{\Delta\chi^2}_{NH}=-\overline{\Delta\chi^2}_{IH}=\overline{\Delta\chi^2}>0$ we have $T_{c}=\mp\overline{\Delta\chi^2}$. One nice feature of this choice is that the CL, expressed as number of $\sigma$'s, takes the well-known form $\sqrt{\overline{\Delta\chi^2}}$. On the other hand, the power for this kind of test is only 0.5, this means there is only 50\% of possibility of getting such a result.
\item The crossing sensitivity, instead, is defined, at least in the symmetric case, taking $T_{c,NH}=T_{c,IH}=0$: the main advantage is that, in this case, the power is equal to the CL (using this criterion, it can be easily defined in a more general scenario); however the number of $\sigma$'s is only $\sqrt{\overline{\Delta\chi^2}}/2$
\end{itemize}

Another common criterion for the discovery is the p-value, which is also defined using the frequentist approach: the difference with the hypothesis test is that while the CL of the latter is defined before the experiment, and the results can only tell if it is achieved or not, the CL of the former depends on the result. The p-value, indeed, is defined as the possibility of finding a ``more extreme'' value of the test statistic than the observed one; in the case of $\Delta\chi^2$, this means $\Delta\chi^2> (<) \Delta\chi^2_{obs}$ if we want to exclude the inverted (normal) hierarchy.  Few remarks about the this approach:
\begin{itemize}
\item It relies on the knowledge of the pdf of $\Delta\chi^2$, however we saw in the previous section that in many cases it can only be determined using Monte Carlo simulations, and it would be difficult to get data reliable at 5 $\sigma$'s or more.
\item Moreover, these distributions depend on the value of other parameters: in particular, in the case of accelerator neutrinos the pdf depends strongly on the value of $\delta_{CP}$. How is it possible to define a CL $1-\alpha$? In \cite{blenFreq}, for the hypothesis test, the authors suggest to define such a CL when for every value of the pull parameters the CL is at least $1-\alpha$, however using this approach it is not clear how to take into account eventual pre-existing constraints (for example, if some values of $\delta_{CP}$ are already excluded at 4 $\sigma$'s, should they still be considered?). Another possible solution (at least for the p-value) is to use the best fit values calculated assuming the hierarchy we want to reject: using the first approach the confidence level could be underestimated, while with the second one it could be overestimated.
\item The frequentist approach can only estimate the compatibility of each hierarchy with the data, namely the CL with which each hierarchy can be excluded; however even if the normal hierarchy can be excluded with a given CL, this does not necessarily means that the hierarchy is inverted. For example, let us assume that the results of an experiment allow us to exclude the normal hierarchy at 5 $\sigma$'s; it would be incorrect, however, to state that ``the hierarchy is determined at 5$\sigma$'s'': indeed, the scenario would be very different if the inverted hierarchy is compatible with the data within 1 $\sigma$ or if it can be excluded at 5 $\sigma$'s, too.
\item Finally, it is worth noticing that, while it is true that using the median frequentist sensitivity the CL takes the familiar form of $\sqrt{\overline{\Delta\chi^2}}$, this is true only for the expected value of $\Delta\chi^2$, not if we use the $\Delta\chi^2$ obtained after we performed the experiment.  Indeed, if we assume that analyzing the result of a certain experiment (or global fit) we found $\Delta\chi^2=16$, it does not follow that the inverted hierarchy can be excluded at 4 $\sigma$'s: the p-value is equal to the median sensitivity only if $\Delta\chi^2=\overline{\Delta\chi^2}$, which in general is not true.
\end{itemize}
\subsection{Bayesian Approach}
While using the frequentist approach it is possible to determine only $P(D|MH)$ (where  $MH=NH,IH$), namely the probability of obtain a certain set of data given the mass hierarchy, using Bayes theorem we can calculate directly $P(MH|D)$ (also called posterior probability), which is the probability for the hierarchy to be normal or inverted given the result of an experiment (these two quantities are deeply different and should not be confused).  $P(MH|D)$ can be obtained using the formula
\begin{equation} \label{bayes}
P(NH|D)=\frac{P(D|NH)\pi(NH)}{P(D|NH)\pi(NH)+P(D|IH)\pi(IH)}=\frac{\pi(NH)}{\pi(NH)+K^{-1}\pi(IH)}
\end{equation}
where $K=P(D|NH)/P(D|IH)=e^{\Delta\chi^2/2}$ is the Bayes factor (in the last step we used Eq. \ref{defChi}), while $\pi(NH/IH)$ are the {\it priors} on the mass hierarchy, namely the degree of belief for some hypothesis (in this case, the mass hierarchies, but priors must be assigned also for all the pull parameters). One of the downside of the Bayesian approach is that the final results depends on the choice of the priors, which are arbitrarily chosen; however in the case of the mass hierarchy there is a very natural solution, namely the symmetric priors, where $\pi(NH)=\pi(IH)=0.5$.

\begin{figure}\begin{center}
\includegraphics[height=3.4cm]{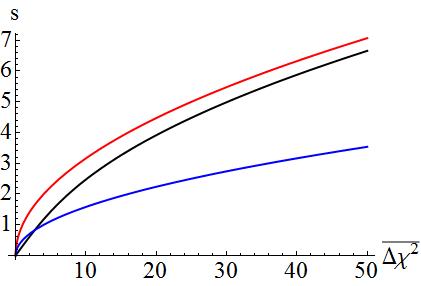}
\includegraphics[height=3.4cm]{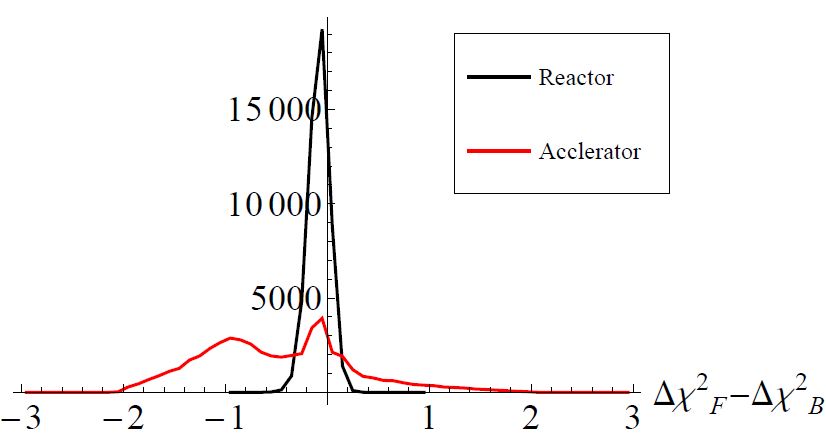}
\end{center}
\caption{\label{comparison}Left Panel: median sensitivity (frequentist: red, Bayesian: black) and crossing sensitivity (blue). Right Panel: Difference between $\Delta\chi^2$ obtained minimizing and marginalizing over the pull parameters}
\end{figure}

In \cite{stat}, the median Bayesian sensitivity is defined as the posterior probability for the mass hierarchy when $\Delta\chi^2=\overline{\Delta\chi^2}$: this probability can be converted into ``number of $\sigma$'s'' using Eq. \ref{sigma}. This definition can be used to quantify the sensitivity of a future experiment, however using the Eq. \ref{bayes} it is possible to calculate this quantity as a function of $\Delta\chi^2$: this means that it can be calculated using only the results of the experiment, while in order to calculate the frequentist CL, one must rely on the knowledge of the pdf of $\Delta\chi^2$. A comparison between the  crossing and the median sensitivity (frequentist and bayesian), as a function of $\overline{\Delta\chi^2}$ can be found in Fig. \ref{comparison} (left panel).

Another advantage of the Bayesian approach is that here all the information can be communicated with a single quantity (the posterior probability) since $P(NH|D)+P(IH|D)=1$ by construction, while using the frequentist approach there is no trivial relation between $P(D|NH)$ and $P(D|IH)$, however it is important to underline that the two methods provide different (and complementary) information: for example, if one experiment can exclude the normal hierarchy at 5$\sigma$'s and the inverted at $3\sigma$'s, while another one at 4 and 1 $\sigma$'s, respectively, the posterior probability would be roughly the same, even though the two scenarios are very different.

Complications may arise if many pull parameters are present: indeed, while in the frequentist approach usually the eventual pull parameters are minimized, in the Bayesian approach we must integrate over them (marginalization), with a weight defined by their priors
\begin{equation}
P(D|MH)=\int \textrm{d}\theta P(D|\theta,MH)\pi(\theta)
\end{equation} 
However these multi-dimensional integrals are usually difficult to compute. If $P(D|\theta,MH)\pi(\theta)$ is strongly peaked around its maximum, and if the determinants of the Hessian matrix for the two hierarchies, calculated in the minima, are the same, it is possible to use the Laplace method to prove that marginalization and minimization yield the same $\Delta\chi^2$ (which can be obtained from $P(D|MH)$ using Eq. \ref{defChi}). This method was applied to the two models described in the previous section, the results are shown in Fig. \ref{comparison} (right panel): we can see that, while in the case of reactor neutrino experiments (at least, in the simplified model considered) this method gives us the correct result to a very good approximation, for accelerator neutrinos this is no longer true: this is to be expected, since we can deduce from Fig. \ref{plotChiAcc} that in case of accelerator neutrinos $P(D|\theta,MH)\pi(\theta)=e^{-\chi^2/2}$ is not peaked around the maximum.
\subsection{Additional Parameter}
A possible way to avoid the non-nested hypotheses issue for the neutrino mass hierarchy was suggested first in \cite{capozzi} for reactor neutrino experiments, where the authors introduced a new, non-physical pull parameter $\eta$, rewriting Eq. \ref{deltaM} as $|\Delta m_{31}^2|=|\Delta m_{32}^2|+(2\eta-1)|\Delta m_{21}^2|$: when $\eta=0$, the hierarchy is inverted, when $\eta=1$ it is normal: in this way one can reduce our problem of model selection to a simpler problem of parameter fitting; however this approach cannot be used, for example, for accelerator neutrinos, since the matter effect depends on the sign of $\Delta m_{31}^2$. A similar but more general approach was suggested in \cite{vanDyk}, where in order to test two hypotheses $H_0$ and $H_1$, which would generate a spectrum $g(x)$ and $f(x)$ respectively, the authors considered the linear combination $g(x)+\eta(f(x)-g(x))$. One advantage of this kind of approach is that the CL for the rejection of both hierarchies can now be expressed in a very compact form, as $\eta\pm\delta\eta$; on the other hand, however, it requires the introduction of a new pull parameter without physical meaning.
\section{Conclusions}
We have presented different approaches, within the frequentist and Bayesian frameworks, for the quantification of the sensitivity in the mass hierarchy determination. While there is no ``correct'' definition and, as long as we are consistent and we specify clearly the convention used, all these approaches can be correct, however it is important to notice that the Bayesian and the frequentist methods give different and complementary information, and it would be preferable to use both, for a more complete analysis.
\Acknowledgements
This work was supported by the National Natural Science Foundation of China (Grants No. 11605247 and 11375201) and by the Chinese Academy of Sciences President's International Fellowship Initiative Grant No. 2015PM063.

\end{document}